\documentclass[12pt]{iopart}

\usepackage[utf8x]{inputenc}
\usepackage{tocloft}
\usepackage{parskip}
\usepackage{float}
\usepackage{upgreek}
\usepackage{multicol}
\usepackage{graphicx}
\usepackage{iopams} 
\begin{document}

\title[Trapping of $^{85}$Rb atoms by metastable state optical pumping]{Trapping of $^{85}$Rb atoms by optical pumping between metastable hyperfine states}

\author{N Cooper and T Freegarde}

\address{School of Physics and Astronomy, University of Southampton, Highfield, Southampton SO17 1BJ, UK}
\ead{nlc2g09@soton.ac.uk}
\begin{abstract}
We describe an atom trapping mechanism based upon differential optical pumping between metastable hyperfine states by partially-displaced laser beams in the absence of a magnetic field. With realistic laser powers, trap spring constants should match or exceed those typical of magneto-optical traps, and highly flexible tailored trap shapes should be achievable. In a proof-of-principle experiment, we have combined a 1D implementation with magneto-optical trapping in the orthogonal directions, capturing $\sim 10^{4}~^{85}$Rb atoms. 

\end{abstract}

\maketitle

\section{Introduction}

Optical atom traps based on the force accompanying the absorption and spontaneous re-emission of radiation are vital tools in atomic physics. The most widely used is the magneto-optical trap (MOT) \cite{raab}, and many experiments can be performed directly on the atom cloud that it forms \cite{motexperiment1,motexperiment2}; when optical dipole or magnetic traps are required, the MOT is usually used for the initial loading into a generally conservative trapping potential \cite{motload1,motload2}. A few alternatives to the MOT have been proposed \cite{spontraps,dalibard,vortex}, each with its advantages and disadvantages. Here, we propose a further alternative, whereby the trap depends upon differential pumping between metastable hyperfine states according to the spatially-varying balance of intensities between different wavelengths of illumination. This metastable optical pumping (MOP) trap may be directly combined with Doppler cooling, but lacks some of the enhanced cooling mechanisms associated with the MOT and is experimentally more complex when implemented in more than one dimension. However, because it depends upon a spatially-varying optical rather than magnetic field, it is more readily sculpted and capable of stronger spring constants. It can in principle operate between the hyperfine states of any atom with a non-zero nuclear spin, or between other states that are sufficiently long-lived.   

The principle underlying our trapping scheme is that if two laser beams address transitions involving two different, long-lived energy levels, then scattering from one beam can be elevated over scattering from the other by optically pumping the atomic population into the appropriate state. This allows the optical Earnshaw theorem \cite{OET} to be circumvented and an optical atom trap to be created.

\begin{figure}
\begin{center}
\includegraphics[scale = 0.315]{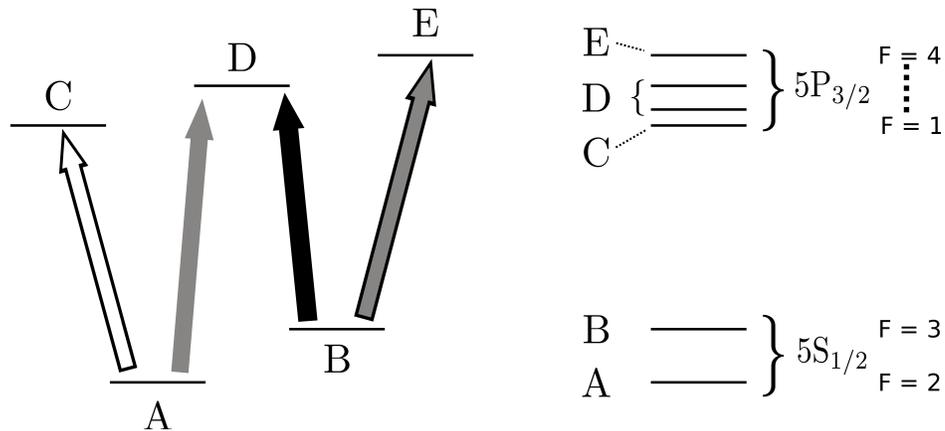}
\caption{Pumping scheme for a MOP trap. The leftmost part shows a generalized level scheme and allowed transitions between the metastable levels A, B and upper radiative levels C, D and E. The corresponding levels of $^{85}$Rb used in our experiments are shown to the right, the 5P$_{3/2}$ F=2 and F=3 states together corresponding to the generalized level D.}
\label{pumpgeom}
\end{center}
\end{figure}

\begin{figure}
\begin{center}
\includegraphics[scale = 0.4]{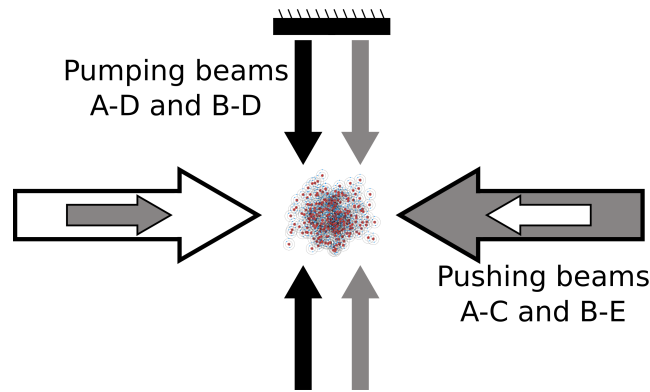}
\caption{Beam geometry for a 1D MOP trap. The transition addressed by each beam is that shown in figure \ref{pumpgeom} by the same arrow colour and border. Larger arrows for the horizontal `pushing' beams indicate higher intensities. The intensity distributions of the vertical `pumping' beams determine the force profile of the trap, although each is balanced against its counter propagating twin to eliminate radiation forces from the pumping beams themselves.}
\label{pumpgeomgeom}
\end{center}
\end{figure}

Figures \ref{pumpgeom} and \ref{pumpgeomgeom} illustrate the key elements in our scheme, which relies upon the spatially-dependent balance between two opposing optical forces, reflecting in turn the spatial dependence of the populations of the metastable levels A and B. Permitted electric dipole transitions exist between states C and A, D and A, D and B, and E and B. Atoms in the metastable state A experience an average scattering force to the right from the power imbalance between the left and right traveling beams tuned to the A-C transition, while atoms in the alternative metastable state B experience an average scattering force in the opposite direction from the imbalance between the beams tuned to the B-E transition. Pump beams A-D and B-D, which propagate across the scattering beams and are individually balanced to exert no average force, provide spatially-dependent transfer of population between the metastable levels. By shaping the pump beams using a mask or spatial light modulator (SLM), the magnitude and direction of the net force upon the atoms may be tailored to the desired trap shape with a near-wavelength resolution that is several orders of magnitude finer than may usually be achieved by shaping the magnetic fields of a magneto-optical trap.

\section{Experimental implementation}

\label{expimp}

In our experimental implementation using atomic $^{85}$Rb, we use the 5S$_{1/2}$ F=2 and F=3 `ground' hyperfine states as the metastable levels A and B respectively and the 5P$_{3/2}$ F=1 to F=4 states as the upper radiative levels C, D and E, as shown in figure \ref{pumpgeom}: the F=2 and F=3 states together correspond to level D and allow favourable branching ratios to be chosen for the respective pumping beams. All our beams are detuned slightly to the red of the relevant atomic transitions to allow them to perform the additional function of Doppler cooling.

For practical convenience, we combine the functions of the A-C rightward pushing beam and the A-D pumping beam by using a single rightward-directed beam tuned to the A-D (5S$_{1/2}$ F=2 $\leftrightarrow$ 5P$_{3/2}$ F=3) transition. In the absence of the other pumping beam (B-D), this concentrates atomic population in the 5S$_{1/2}$ F=3 ground hyperfine state B, where atoms experience only the leftward force exerted by the pushing beam pair B-E, which is tuned to the 5S$_{1/2}$ F=3 $\leftrightarrow$ 5P$_{3/2}$ F=4 transition and which experimentally corresponds to a bright beam incident from the right and its attenuated reflection. Where the atoms are also illuminated by the pumping beam B-D, tuned to the 5S$_{1/2}$ F=3 $\leftrightarrow$ 5P$_{3/2}$ F=2 transition, atoms may be pumped into the ground hyperfine state A, where they experience a rightward force due to the A-D beam that can overcome that due to scattering of the B-E pushing beam. The net force upon the atoms depends upon the balance between the metastable state populations and hence upon the local intensity of the B-D pumping beam: where this is high there is a net force to the right, where it is weak the net force is to the left, and the trap centre forms where the net force is zero on the right-hand edge of the B-D pumping beam.

\begin{figure}
\begin{center}
\includegraphics[scale=1.50]{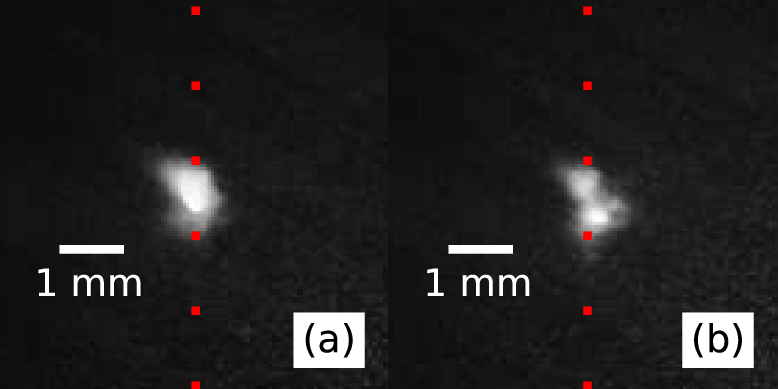}
\caption{Images of the atom clouds formed by a 1D MOP trap combined with 2D magneto-optical trapping in the orthogonal directions. In (b) a thin wire was placed across the centre of the B-D beam, casting a geometric shadow into the trapping region that split the B-D beam in two and thus generated two adjacent trapping sites separated by a small area of reduced atom density. The dotted red lines show the nominal axis of the MOP trap.}
\label{dots}
\end{center}
\end{figure}

As a proof-of-principle experiment, we combined a two-dimensional magneto-optical trap, usually used to bring an atom cloud close to a microstructured surface, with a one-dimensional MOP trap in the orthogonal direction. All beams along the trapping axis of the MOP trap were linearly polarised to eliminate magneto-optical forces in this direction. Around 10$^4$ $^{85}$Rb atoms were captured - about a third of the number obtained with a full 3-D MOT in this configuration \cite{lambdamot}. The trapped atom cloud is shown in figure \ref{dots}(a). To demonstrate the potential for sculpting of the trap geometry, a thin wire was placed across the centre of the B-D beam, casting a geometric shadow into the trapping region and thereby splitting the B-D beam in two, thus creating two adjacent trapping regions as shown in figure \ref{dots}(b).

\begin{figure}
\begin{center}
\includegraphics[scale=0.8]{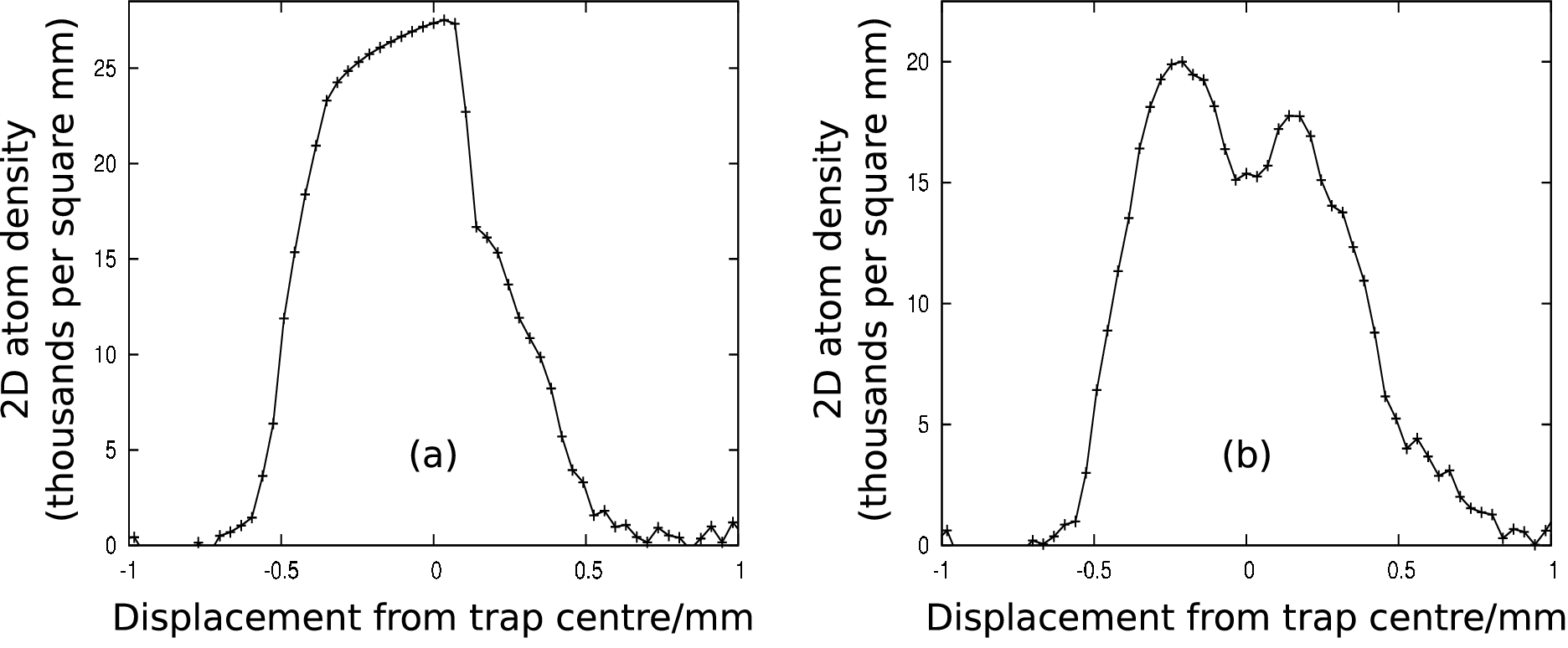}
\caption{Two-dimensional atom density as a function of position along the axis of the MOP trap, for the atom clouds shown in figure \ref{dots}. In (b) a thin wire was placed across the centre of the B-D beam, casting a geometric shadow into the trapping region that split the B-D beam in two and thus generated two adjacent trapping sites separated by a small area of reduced atom density.}
\label{denplots}
\end{center}
\end{figure}

The axis of the MOP trap was nominally along the vertical direction in figure \ref{dots}, where it is shown by the dotted red lines, and figure \ref{denplots} gives quantitative data for the two-dimensional atom density as a function of position along the axis of the MOP trap both with and without obstruction of the B-D beam. As the depth of field of our imaging system was large compared with the dimensions of the atom cloud, the measured atom density is integrated along the direction normal to the plane of the images in figure \ref{dots}. The extent of the atom cloud was slightly greater along the axis of the MOP trap than in the orthogonal direction, but this results from the specific conditions used in our experimental implementation and is not reflective of any fundamental limitation.
  
By releasing and subsequently imaging the trapped atoms, the temperature of the cloud in the hybrid trap was found to be 93$\pm$9~$\mu$K. This process also revealed that the one-dimensional velocity distribution along the axis of the MOP trap was not significantly different from that along the orthogonal axis. When magneto-optical trapping was employed in all three dimensions, the temperature of the resulting cloud was found to be 110$\pm$40~$\mu$K \cite{lambdamot}. Both of these results are consistent with our expectation that a MOP trap should exhibit lower levels of sub-Doppler cooling than a conventional MOT, as the same is true of the non-standard MOT employed in this work. We would therefore expect a full, three-dimensional MOP trap to produce an atom cloud with a temperature similar to that measured in our hybrid trap.    

In addition to ensuring that all beams propagating along the axis of the MOP trap were linearly polarised, we carried out a control experiment to verify that our results could not be explained by residual magneto-optical trapping. If atom trapping had been occurring via a mechanism that was not dependent on optical pumping, then it should have been possible to obtain similar results when the B-D and/or A-D beams were replaced with beams addressing the B-E transition, provided that the optical power in these beams was correctly adjusted so as to restore the original balance of radiation pressure forces in the trapping region. This experiment was carried out and, despite careful observation during an extensive scan of the beam powers, no trapping was seen when either the B-D or A-D beam was replaced with light addressing the B-E transition. To ensure that the replacement beams occupied the same spatial modes as the original beams, they were conveyed to the experiment through the same single-mode optical fibers, with no change in the alignment of the output couplers.

\section{Theoretical Analysis}

\begin{figure}
\begin{center}
\includegraphics[scale=1.5]{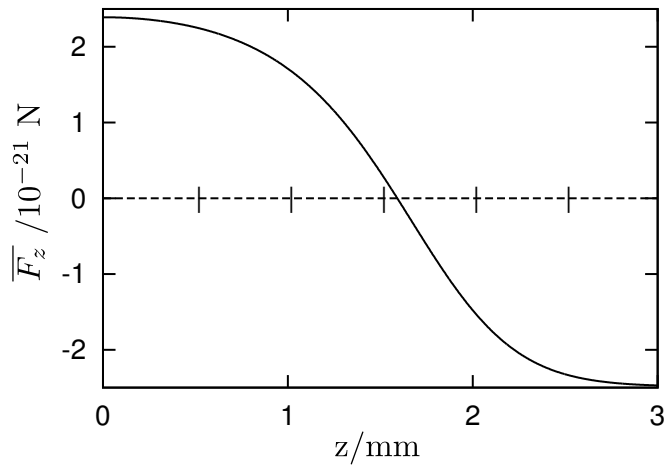}
\caption{Time-averaged z component of the radiation pressure force on an atom as a function of its position on the z axis for a one dimensional MOP trap based on the scheme shown in figures \ref{pumpgeom} and \ref{pumpgeomgeom} and the parameters given in (9--12). The origin is taken to be on the axis of the B-D beam.}
\label{zforce}
\end{center}
\end{figure}

To gain an approximate quantitative understanding of the mechanism, we model the atoms as five-level systems governed by a set of rate equations, with stimulated and spontaneous transition rates between, for example, states E and B given by $\tau_{\mathrm{EB}}I_{\mathrm{EB}}$ and $\Gamma_{\mathrm{EB}}$ respectively, where $I_{\mathrm{EB}}$ is the intensity of the laser light tuned to the relevant transition. Transition rates between other pairs of states are labeled similarly. The five relevant levels in our experimental implementation are A, B, E and the two available D states, which we shall label D$_2$ and D$_3$ according to their angular momentum state F = 2 or F = 3 in the subsequent discussion. We ignore the coupling of the beams to non-target transitions and neglect the atomic coherences which, if the lasers are detuned by the order of a linewidth for Doppler cooling, are unlikely to play a significant role --- theoretical models that ignore atomic coherences have been shown to accurately predict experimental results under similar circumstances \cite{mattpump}. As the timescales associated with motion within the trap are much longer than those associated with optical pumping, we use the steady-state populations and time-averaged forces to determine the trap properties. We derive the stimulated transition rates by considering the steady-state solution of the optical Bloch equations for a two-level system characterised by a spontaneous decay rate $\Gamma$, when illuminated by a beam of detuning $\delta$ and intensity $I$, with Rabi frequency $\Omega$, and comparing this to the steady-state populations in a simple rate equation model. Equating the results for the upper state population, we obtain

\begin{equation}
\frac{\Omega^{2}/4}{\delta^{2}+\Omega^{2}/2+\Gamma^{2}/4} = \frac{\tau I}{2\tau I+\Gamma},
\end{equation}

and therefore, defining the dipole matrix element $\langle \mathrm{E}|x|\mathrm{B} \rangle$ between two levels as $X_{\mathrm{EB}}$, we obtain

\begin{equation}
\tau_{\mathrm{EB}} = \frac{\Omega_{\mathrm{EB}}^{2} \Gamma_{\mathrm{EB}}}{4 I_{\mathrm{EB}}(\delta_{\mathrm{EB}}^{2}+\Gamma_{\mathrm{EB}}^{2}/2)} = \frac{e^{2} \left| X_{\mathrm{EB}} \right|^{2} \Gamma_{\mathrm{EB}}}{2 \hbar^{2} c \epsilon_{0} (\delta_{\mathrm{EB}}^{2}+\Gamma_{\mathrm{EB}}^{2}/2)}
\end{equation}

and similar expressions for the other transitions. The rate equations governing the system are then:

\begin{equation}
\frac{\mathrm{dE}}{\mathrm{dt}} = (\mathrm{B}-\mathrm{E})\tau_{\mathrm{EB}}I_{\mathrm{EB}}-\mathrm{E}\Gamma_{\mathrm{EB}},
\end{equation}

\begin{equation}
\frac{\mathrm{dD}_{3}}{\mathrm{dt}} = (\mathrm{A}-\mathrm{D}_{3})\tau_{\mathrm{D}_{3}\mathrm{A}}I_{\mathrm{D}_{3}\mathrm{A}}-\mathrm{D}_{3}\Gamma_{\mathrm{D}_{3}\mathrm{B}}-\mathrm{D}_{3}\Gamma_{\mathrm{D}_{3}\mathrm{A}},
\end{equation}

\begin{equation}
\frac{\mathrm{dD}_{2}}{\mathrm{dt}} = (\mathrm{B}-\mathrm{D}_{2})\tau_{\mathrm{D}_{2}\mathrm{B}}I_{\mathrm{D}_{2}\mathrm{B}}-\mathrm{D}_{2}\Gamma_{\mathrm{D}_{2}\mathrm{B}}-\mathrm{D}_{2}\Gamma_{\mathrm{D}_{2}\mathrm{A}},
\end{equation}

\begin{equation}
 \frac{\mathrm{dB}}{\mathrm{dt}} = (\mathrm{E}-\mathrm{B})\tau_{\mathrm{EB}}I_{\mathrm{EB}}+(\mathrm{D}_{2}-\mathrm{B})\tau_{\mathrm{D}_{2}\mathrm{B}}I_{\mathrm{D}_{2}\mathrm{B}}+\mathrm{E}\Gamma_{\mathrm{EB}}+\mathrm{D}_{3}\Gamma_{\mathrm{D}_{3}\mathrm{B}}+\mathrm{D}_{2}\Gamma_{\mathrm{D}_{2}\mathrm{B}},
\end{equation}

and

\begin{equation}
\frac{\mathrm{dA}}{\mathrm{dt}} = (\mathrm{D}_{3}-\mathrm{A})\tau_{\mathrm{D}_{3}\mathrm{A}}I_{\mathrm{D}_{3}\mathrm{A}}+\mathrm{D}_{3}\Gamma_{\mathrm{D}_{3}\mathrm{A}}+\mathrm{D}_{2}\Gamma_{\mathrm{D}_{2}\mathrm{A}},
\end{equation}

where A-E represent the populations of the states A-E shown in figure \ref{pumpgeom}. Setting all the time derivatives to zero and solving these equations yields the steady-state populations of the five levels, as described in the Appendix, from which the time-averaged force on an atom may be found to be

\begin{equation}
\fl \overline{\mathbf{F}} = (\mathrm{B}_{s}-\mathrm{E}_{s})\tau_{\mathrm{EB}}I_{\mathrm{EB}}\mathbf{p}_{\mathrm{EB}} + (\mathrm{A}_{s}-\mathrm{D}_{3s})\tau_{\mathrm{D}_{3}\mathrm{A}}I_{\mathrm{D}_{3}\mathrm{A}}\mathbf{p}_{\mathrm{D}_{3}\mathrm{A}} + (\mathrm{B}_{s}-\mathrm{D}_{2s})\tau_{\mathrm{D}_{2}\mathrm{B}}I_{\mathrm{D}_{2}\mathrm{B}}\mathbf{p}_{\mathrm{D}_{2}\mathrm{B}},
\end{equation}

where $\mathbf{p}_{\mathrm{EB}}$ etc. are the (vector) mean photon momenta associated with scattering of the corresponding laser beams and the subscript $s$ indicates the steady-state populations. Substitution of some realistic experimental parameters allows comparison of the restoring force and its gradient in the MOP trap with values typical of a MOT. For the one-dimensional trap of figure \ref{pumpgeom}, with 10~mW each in 3~mm diameter beams and 30\% attenuation of the B-E beam on reflection, detunings of $\sim$2$\Gamma$, and with the pushing beams propagating along the z-axis, we hence have beam parameters

\begin{equation}
I_{\mathrm{EB}}(x,y,z) = 1.7 I_{0} \mathrm{exp}\left[-2(x^{2}+y^{2})/r_{0}^{2} \right], 
\end{equation}

\begin{equation}
I_{\mathrm{D}_{3}\mathrm{A}}(x,y,z) = I_{0} \mathrm{exp}\left[-2(x^{2}+y^{2})/r_{0}^{2} \right], 
\end{equation}

\begin{equation}
I_{\mathrm{D}_{2}\mathrm{B}}(x,y,z) = 2 I_{0} \mathrm{exp}\left[-2(x^{2}+z^{2})/r_{0}^{2} \right],
\end{equation}

\begin{equation}
 \mathbf{p}_{\mathrm{EB}} = \frac{-0.18h}{\lambda}\hat{\mathbf{z}},~~~~~~ \mathbf{p}_{\mathrm{D}_{3}\mathrm{A}} = \frac{h}{\lambda}\hat{\mathbf{z}} ~~~~~~ \& ~~~~~~  \mathbf{p}_{\mathrm{D}_{2}\mathrm{B}} = \mathbf{0},
\end{equation}

where $I_0 = 2.8\times10^3$~Wm$^{-2}$ and $r_0 = 1.5$~mm. Using the $^{85}$Rb parameters from \cite{steck} and assuming the trapping light to be unpolarised, we hence obtain the trapping force profile shown in figure \ref{zforce}, which shows the trap centre to be about 1.6~mm from the axis of the B-D laser beam and the trap to have a spring constant around 4$\times10^{-18}$~Nm$^{-1}$. This is one or two orders of magnitude greater than the spring constant in a typical MOT, which is proportional to the magnetic field gradient about the trap centre and is therefore constrained by practical rather than fundamental considerations, usually to something of the order of 10$^{-19}$~Nm$^{-1}$ or below \cite{motspringconsts}.

\begin{figure}
\begin{center}
\includegraphics[scale=0.6]{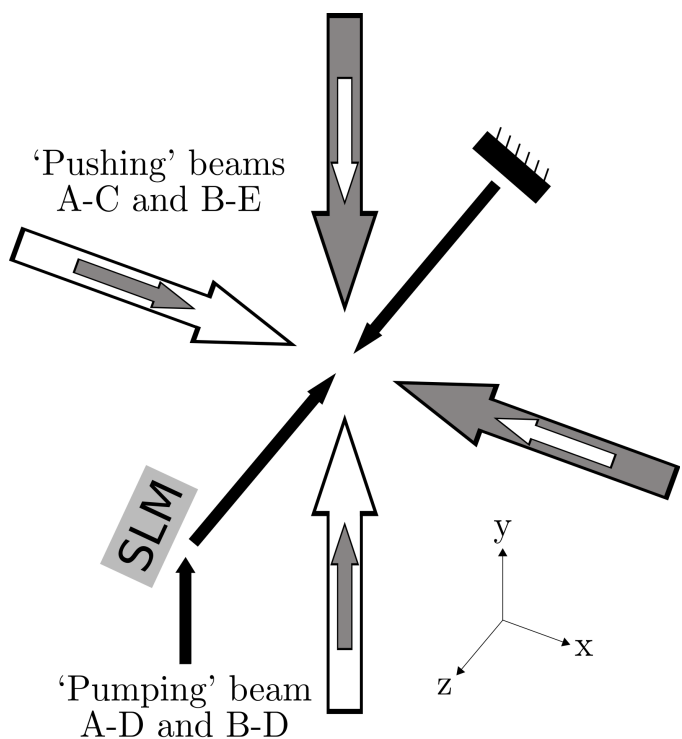}
\caption{Proposed beam geometry for 2D trapping via time-multiplexing. The pairs of opposing `pushing' beams would only be active along either the x or y axis at any given time, with switching between these beams being achieved, for example, via the use of a Pockels cell and polarising beam splitter. The spatial light modulator (SLM) would be used to switch the intensity profile of the `pumping' beam, with simultaneous sculpting of the A-D and B-D beams being possible through, for example, the use of distinct spatial regions of the SLM.}
\label{2dtrap}
\end{center}
\end{figure}

The trap described thus far operates in just one dimension, as the balance of forces within different pushing beam pairs can only be independently modified if the pushing beam pairs address distinct pairs of metastable levels. Three dimensional trapping would therefore be possible if a third metastable level were included in the pumping scheme. However, the most straightforward extension to two and three dimensions will be by time-multiplexing, as illustrated schematically in figure \ref{2dtrap}, whereby independent one-dimensional traps operate along orthogonal axes, with high frequency switching of the laser beams to alternate between the separate traps. The intensity profile of the pumping beams in two or three dimensional traps could be set using a single 2D SLM if this switching were synchronized with the SLM frame-update frequency, which would also facilitate dynamic adjustment of the trap shape. A similar time-multiplexing scheme, in which alternation between two 2D magneto-optical traps results in a full, 3D MOT, has recently been demonstrated experimentally \cite{DMOT}. It may also be possible to generate an interesting range of trapping geometries by using Laguerre-Gaussian or other `exotic' beam modes \cite{reviewer2} for the pumping beams.

\section{Applications and Conclusions}

One potential application of the MOP trap is to experiments where a particular atom density distribution is required, for example for the efficient loading of an array of conservative atom traps of the type discussed in \cite{microlens,atomchip}. It might also be helpful to combine the large spring constant and highly flexible shape of the MOP trap with the efficient cooling and loading of a MOT in order to obtain a very high density of trapped atoms. A significant advantage of the MOP trap, however, is that it requires no magnetic field or field gradient, and is therefore compatible with trapping near to magnetic materials and structures that would disturb a conventional magneto-optical trap. The absence of a magnetic component also allows the trapping potential to be modified more rapidly than in a MOT, where inductive effects usually place a practical limit on the rate at which the magnetic field can be altered. Furthermore, as the trapping mechanism does not rely upon population transfer between Zeeman sublevels, MOP traps share with those of \cite{vortex} the option to spin-polarise the atom cloud to some degree if required.

Zeeman-assisted slowing and Sisyphus cooling are not inherently present in a MOP trap as they are in a MOT; this means that a MOP trap is likely to produce a smaller atom cloud with a higher temperature than that generated by a typical MOT. The experimental complexity is also greater for multi-dimensional traps. However, as the optical polarization plays no role in the MOP trapping mechanism, there is no reason why, in the presence of an additional magnetic field, the beam polarizations could not be chosen so as to promote sub-Doppler cooling for lower temperatures and higher loading rates.

\section{Acknowledgements}

This work was supported by the UK EPSRC grants EP/E039839/1 and EP/E058949/1. The authors would like to thank James Bateman for helpful discussions.

\appendix
\section*{Appendix}
\setcounter{section}{1}

Given here are the full steady-state solutions to the rate equations for the five-level system. In order to simplify the formulae we introduce the following notation:

\begin{equation}
K_{1} = \frac{\tau_{\mathrm{EB}}I_{\mathrm{EB}}}{\tau_{\mathrm{EB}}I_{\mathrm{EB}} + \Gamma_{\mathrm{EB}}},
\end{equation}

\begin{equation}
K_{2} = \frac{\tau_{\mathrm{D}_{3}\mathrm{A}}I_{\mathrm{D}_{3}\mathrm{A}}}{\tau_{\mathrm{D}_{3}\mathrm{A}}I_{\mathrm{D}_{3}\mathrm{A}} + \Gamma_{\mathrm{D}_{3}\mathrm{B}} + \Gamma_{\mathrm{D}_{3}\mathrm{A}}},
\end{equation}

\begin{equation}
K_{3} = \frac{\tau_{\mathrm{D}_{2}\mathrm{B}}I_{\mathrm{D}_{2}\mathrm{B}}}{\tau_{\mathrm{D}_{2}\mathrm{B}}I_{\mathrm{D}_{2}\mathrm{B}} + \Gamma_{\mathrm{D}_{2}\mathrm{B}} + \Gamma_{\mathrm{D}_{2}\mathrm{A}}},
\end{equation}

and

\begin{equation}
K_{4} = \frac{-K_{3}\Gamma_{\mathrm{D}_{2}\mathrm{A}}}{\Gamma_{\mathrm{D}_{3}\mathrm{A}}K_{2} + (K_{2}-1)\tau_{\mathrm{D}_{3}\mathrm{A}}I_{\mathrm{D}_{3}\mathrm{A}}}.
\end{equation}

For the steady state populations of the five levels we then obtain:

\begin{equation}
\mathrm{B}_{s} = (1+K_{1}+K_{2}K_{4}+K_{3}+K_{4})^{-1},
\end{equation}

\begin{equation}
\mathrm{E}_{s} = K_{1}\mathrm{B}_{s},
\end{equation}

\begin{equation}
\mathrm{D}_{3s} = K_{2}K_{4}\mathrm{B}_{s},
\end{equation}

\begin{equation}
\mathrm{D}_{2s} = K_{3}\mathrm{B}_{s},
\end{equation}

and

\begin{equation}
\mathrm{A}_{s} = K_{4}\mathrm{B}_{s}.
\end{equation}

Combining these with (8)--(12) allows the derivation of a full expression for the (time-averaged) force as a function of position, $\overline{\mathbf{F}}(x,y,z)$. Once this is established, the trap centre can be found by setting $\overline{\mathbf{F}}(x,y,z) = \mathbf{0}$ and solving for the coordinates at which this equality holds. It is then possible to calculate the spring constant along a given axis by computing $\left. \frac{\mathrm{d}\overline{F}_{i}}{\mathrm{d}x_{i}}\right|_{\mathrm{trap~centre}}$.


\section*{References}
\begin{thebibliography}{15}

\bibitem{raab} Raab E, Prentiss M, Cable A, Chu S and Pritchard D 1987 Phys. Rev. Lett. \textbf{59} 2631 
\bibitem{motexperiment1} Gabbanini C, Fioretti A, Lucchesini A, Gozzini S and Mazzoni M 2000 Phys. Rev. Lett. \textbf{84} 2814 
\bibitem{motexperiment2} Li W, Mourachko I, Noel M and Gallagher T 2003  Phys. Rev. A \textbf{67} 052502 
\bibitem{motload1} Roy A, Jing A and Barrett M 2012 New J. Phys. \textbf{14} 093007 
\bibitem{motload2} Viteau M, Radogostowicz J, Chotia A, Bason M, Malossi N, Fuso F, Ciampini D, Morsch O, Ryabtsev I and Arimondo E 2010 J. Phys. B \textbf{43} 155301 
\bibitem{spontraps} Pritchard D, Raab E, Bagnato V, Wieman C and Watts R 1986 Phys. Rev. Lett. \textbf{57} 310-313
\bibitem{dalibard} Bouyer P, Lemonde P, Dahan M, Michaud A, Salomon C and Dalibard J 1994 Europhys. Lett. \textbf{27} 569 
\bibitem{vortex} Walker T, Feng P, Hoffmann D and Williamson R 1992 Phys. Rev. Lett. \textbf{69} 2168 
\bibitem{OET} Ashkin A and Gordon J 1983 Opt. Lett. \textbf{8} 511-513
\bibitem{lambdamot} Ohadi H, Himsworth M, Xuereb A and Freegarde T 2009 Opt. Expr. \textbf{17} 23003 
\bibitem{mattpump} Himsworth M and Freegarde T 2010 Phys. Rev. A \textbf{81} 023423
\bibitem{steck} Steck D 2012 http://steck.us/alkalidata (revision 2.1.5)
\bibitem{motspringconsts} Kim K, Noh H and Jhe W 2005 Phys. Rev. A \textbf{71} 033413
\bibitem{DMOT} Rushton J, Bateman J, Aldous M and Himsworth M (in preparation) `A Dynamic Magneto-Optical Trap for Atom Chips'


\bibitem{reviewer2} Arlt J, Dholakia K, Allen L and Padgett M 1998 J. Mod. Opt. \textbf{45} 1231-1237 



\bibitem{microlens} Kruse J, Gierl C, Schlosser M, and Birkl G 2010 Phys. Rev. A \textbf{81} 060308(R) 
\bibitem{atomchip} Fort\'{a}gh J and Zimmermann C 2007 Rev. Mod. Phys. \textbf{79} 235-289


\end{thebibliography}
\end{document}